\documentclass[journal=ancac3,manuscript=article, layout=twocolumn]{achemso}

\usepackage[version=3]{mhchem} 
\usepackage{siunitx}
\usepackage{graphicx}
\usepackage[label font=bf,labelformat=simple]{subfig}
\usepackage{caption}
\usepackage{chemformula}

\newcommand*\nc{NC}
\newcommand*\ncs{NCs}

\newcommand*\of[1]{\left(#1\right)}

\author{Stefano Pierini}
\affiliation[UTT]{Laboratory Light, nanomaterials \& nanotechnologies – L2n, University of Technology of Troyes \& CNRS ERL 7004, 12 rue Marie Curie, 10000 Troyes, France}
\alsoaffiliation[LKB]{Laboratoire Kastler Brossel, Sorbonne Universit\'e, CNRS, ENS-PSL Research University, Coll\`ege de France, 4 place Jussieu, 75252 Paris Cedex 05, France}

\author{Marianna D'Amato}
\affiliation{Laboratoire Kastler Brossel, Sorbonne Universit\'e, CNRS, ENS-PSL Research University, Coll\`ege de France, 4 place Jussieu, 75252 Paris Cedex 05, France}

\author{Mayank Goyal}
\affiliation[INSP]{Sorbonne Université, CNRS - UMR 7588, Institut des NanoSciences de Paris, INSP, F-75005 Paris, France}
\alsoaffiliation{Department of Chemistry, Indian Institute of Science Education and Research (IISER), Pune 411008, India}

\author{Quentin Glorieux}
\affiliation{Laboratoire Kastler Brossel, Sorbonne Universit\'e, CNRS, ENS-PSL Research University, Coll\`ege de France, 4 place Jussieu, 75252 Paris Cedex 05, France}

\author{Elisabeth Giacobino}
\affiliation{Laboratoire Kastler Brossel, Sorbonne Universit\'e, CNRS, ENS-PSL Research University, Coll\`ege de France, 4 place Jussieu, 75252 Paris Cedex 05, France}

\author{Emmanuel Lhuillier}
\affiliation[INSP]{Sorbonne Université, CNRS - UMR 7588, Institut des NanoSciences de Paris, INSP, F-75005 Paris, France}

\author{Christophe Couteau}
\affiliation{Laboratory Light, nanomaterials \& nanotechnologies – L2n, University of Technology of Troyes \& CNRS ERL 7004, 12 rue Marie Curie, 10000 Troyes, France}

\author{Alberto Bramati}
\affiliation{Laboratoire Kastler Brossel, Sorbonne Universit\'e, CNRS, ENS-PSL Research University, Coll\`ege de France, 4 place Jussieu, 75252 Paris Cedex 05, France}
\email{alberto.bramati@lkb.upmc.fr}

\title[]
  {Highly photo-stable Perovskite nanocubes: towards integrated single photon sources based on tapered nanofibers
  }


\begin{document}
\begin{abstract}

The interest in perovskite nanocrystals (\ncs{}) such as \ch{CsPbBr3} for quantum 
applications is rapidly raising, as it has been demonstrated that they can behave as very efficient single 
photon emitters. The main problem to tackle in this context is their photo-stability under optical excitation. In this article, we present a full analysis of the optical and quantum properties of highly efficient perovskite nanocubes synthesized with an established method, which is used for the first time to produce quantum emitters, and is shown to ensure an increased photo-stability. These emitters exhibit reduced blinking together with a strong photon antibunching. Remarkably these features are hardly affected by the increase of the excitation intensity well above the emission saturation levels. 
Finally, we achieve for the first time the coupling of a single perovskite nanocube with a
tapered optical nanofiber in order to aim for a compact integrated single photon source for future applications.
\end{abstract}

\section{}
The interest in perovskites, originally studied for solar-cell applications~\cite{park2015perovskite}, has recently increased in the quantum optics community. Perovskite nanocrystals are indeed versatile emitters and the possibility to tune their emission wavelength playing on their size and composition, together with their coherent emission\cite{utzat2019Coherent} and their ability to obtain single photon emission at low~\cite{rainoSingleCesiumLead2016,huo2020Optical} and room~\cite{parkRoomTemperatureSinglePhoton2015} temperatures makes them promising nano-objects for quantum applications.
Despite these strong points\cite{raino2018Superfluorescence}, the optical stability is still the main limitation in their use as they usually bleach after few minutes under illumination. Multiple approaches have been attempted to reduce this problem such as polymer encapsulation~\cite{chenInfluencePMMAAllInorganic2019,rainoUnderestimatedEffectPolymer2019}, but also alumina encapsulation using atomic layer deposition\cite{saris2019Optimizing,xiang2018Bottom}, and surface passivation~\cite{panAirStableSurfacePassivatedPerovskite2015}. Although these methods have shown some effects in reducing the bleaching, only partial results are achieved. Indeed, the first technique is not suitable for applications such as the coupling of single emitters with photonic devices, as it requires the \ncs{} to be surrounded by a dense polymer matrix. The second technique is more promising, however no single photon emission has been demonstrated so far using this method.

We report here a method of fabrication, used for the first time for this purpose, that allows us to obtain higher stability samples of perovskite nanocrystals which can be excited under optical excitation for more than one hour. We also investigate the role of the dilution on the stability, suggesting new approaches to address this problem.
Thanks to this improved stability, we were able to perform a full characterization of the optical and quantum features of perovskite nanocrystals, showing at the same time reduced blinking and strong photon antibunching of the emission.
Finally, for the first time with such kind of emitters,  we achieve  the coupling of a single perovskite nanocrystal with a tapered optical nanofiber. As it has been shown with atoms\cite{le2005spontaneous,klimov2004spontaneous} and solid state emitters\cite{fujiwara2011highly,yalla2012fluorescence,joos2018polarization,vorobyovCouplingSingleNV2016} this technique is of paramount importance for applications in the emerging field of quantum technologies.  Our result is a promising step towards the realization of a compact integrated single photon device at room temperature with perovskite nanoemitters. 

\section{Perovskite nanocubes fabrication}

\begin{figure}
    \centering
    \includegraphics[width=\columnwidth]{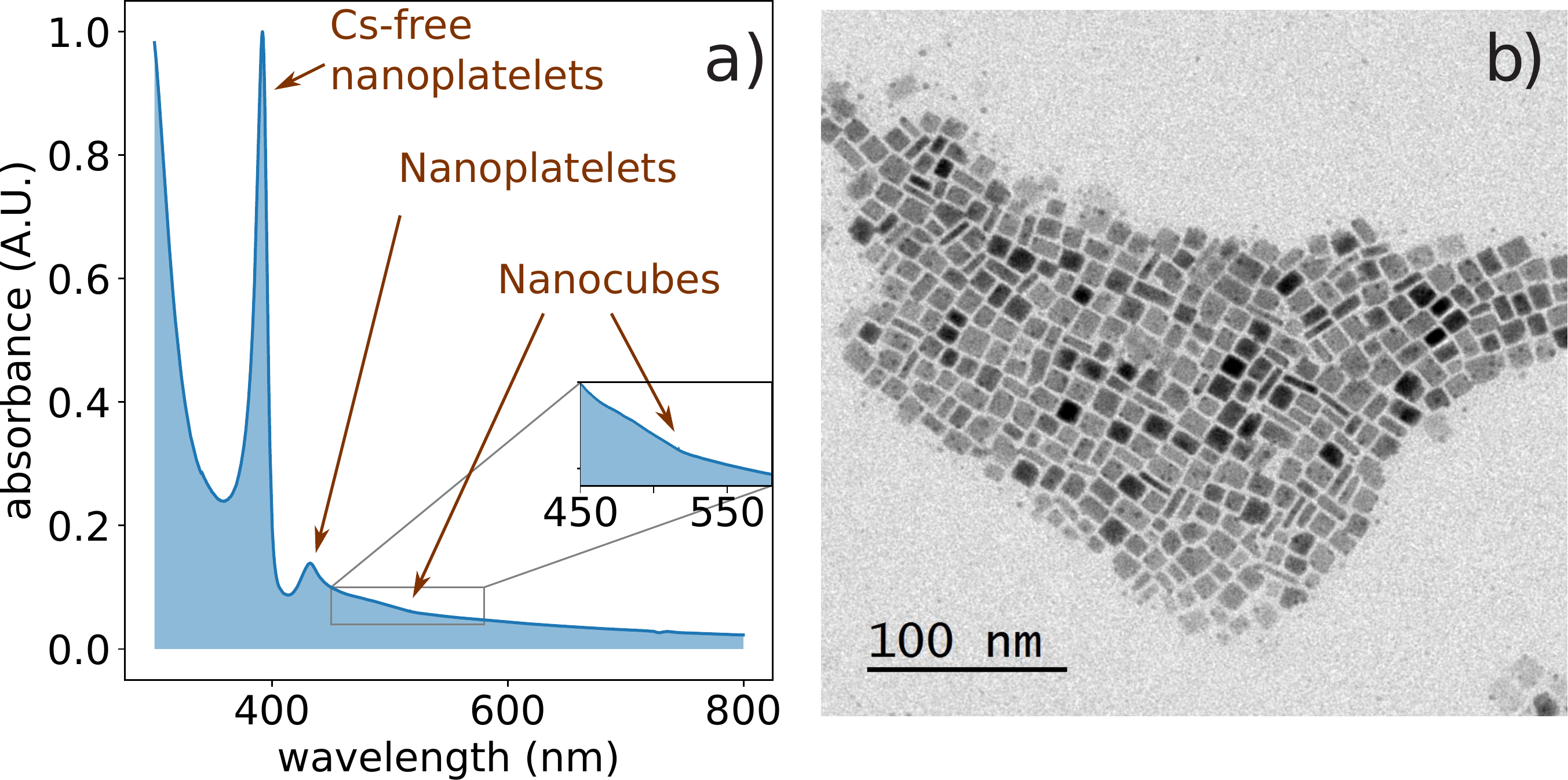}
    \caption{
      a)~Absorption spectrum of the products obtained from the synthesis with the method described in the text before the last centrifugation. The contribution of the three phases is evidenced. The inset shows the portion of the spectrum corresponding to nanocubes contribution.
       b)~Transmission electron microscopy  image of \ch{CsPbBr3} nanocubes responsible for the photoluminescence.
        }
    \label{fig:lhu}
\end{figure}

We start by describing the growth method of nanocrystals of \ch{CsPbBr3} presenting a bright green photoluminescence (PL) (see Figure~\ref{fig:spettri}a) around \SI{500}{\nano \meter}.
The usual procedure as decribed by Protecescu \emph{et al.}\cite{protesescu2015nanocrystals} leads to bright nanoparticles, however
they appear to be difficult to use for single emitter photonic applications due to a limited
colloidal stability under dilute condition. The alkyl ammonium ligands have strong binding dynamics~\cite{de2016highly} to the nanocrystals surface which reduce their stability under dilute condition.
 It was recently proposed to use alternative ligands such as zwitterion~\cite{krieg2019stable} or phosphonic acid\cite{zhang2019alkyl} 
 to enhance the binding of the ligand and overcome the loss of stability under dilution.
 Here, we explore a different approach to obtain nanocrystals more stable under dilute
 condition. We use a procedure initially developed for the growth of \ch{CsPbBr3} nanosheet.\cite{liu2019self} Compared to Protecescu \emph{et al.}\cite{protesescu2015nanocrystals} 
 there are three major changes:
 \begin{enumerate}
     \item less caesium oleate is introduced to favor the growth of Cs free phase.
     \item two additional ligands with saturated alkyl chains are introduced (octanoic acid and octyl amine) to favor the crystallization of the Cs free phase.
     \item the reaction time is extended from \SI{30}{\second} to \SI{35}{min} to favor the reaction step.
 \end{enumerate}
 In this synthesis, the main product is \ch{CsPbBr3} nanoplatelets~\cite{weidman2016highly} and presents an absorption peak at \SI{430}{\nano \meter} as shown in Figure~\ref{fig:lhu}a. Moreover, there are two additional products, the first one is made of \ch{Cs}-free nanoplatelets where a plane of lead bromide is sandwiched between two planes of ligands with C\,8 chains.
 This phase presents a clear peak at \SI{398}{\nano \meter} in the absorption spectrum\cite{weidman2016highly}. 
 The second one is composed of synthesized cubes responsible for the small absorption edge around \SI{510}{\nano \meter}, and it is actually the one we are interested in. We can see its signature as a small change in the slope of the absorption curve, visible in the inset of Figure~\ref{fig:lhu}a.

It turns out that these nanocubes, despite the fact that they are a side product of the synthesis, are very interesting for single photon emitters as we will show. 
In the following, we select only the PL from these cubes. Almost all the \ch{CsPbBr3} nanoplatelets are removed by centrifuging the solution as explained in methods. The obtained \ch{CsPbBr3} nanocubes appear to be more stable upon 
dilution that the one obtained by the direct procedure, as shown in the following.

\subsection{Optical characterization}
\begin{figure}[ht]
  \includegraphics[width=\linewidth]{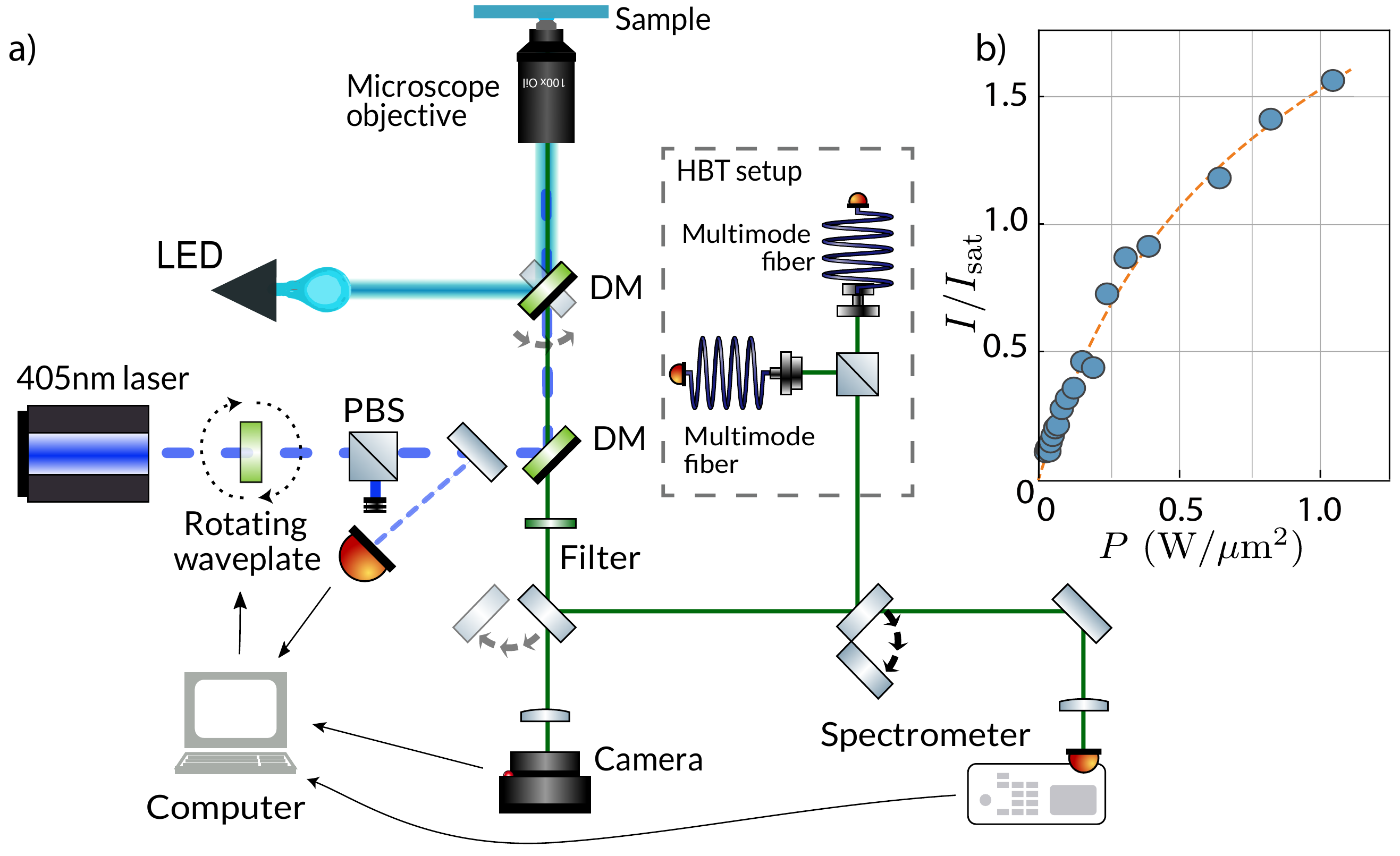}
  \caption{a)~Inverted confocal microscope setup for the sample analysis. The emitters can be excited with a LED or a pulsed laser. The emission can be detected on a camera or analyzed with a spectrometer. Finally, we can perform a $g^{(2)}$ measurement with a Hanbury Brown and Twiss setup (HBT setup), with multimode optical fibers connected to APDs. DM: dichroic mirror, PBS: polarizing beam splitter.
  b)~Saturation measurement of a single \nc{}. The blue dots are the experimental data while the orange line is the fitted function from equation~\protect\eqref{eq:saturation}.%
  }
  \label{sch:setup1}
\end{figure}

We optically characterized the \ncs{} using an inverted confocal microscope: 
the experimental setup is shown in Figure~\ref{sch:setup1}a.

The samples are prepared for optical characterization measurements by spin-coating the colloidal solution on a glass coverslip.
A LED lamp at \SI{400}{\nano \meter} is first used to locate the emitters and an image of the field 
of view of the microscope is collected by a CMOS camera. 
A single emitter is then excited via a \SI{405}{\nano \meter} picosecond
pulsed laser (pulse width $<\SI{50}{ps}$) with a repetition rate adjustable from \SI{2.5}{\mega \hertz} to \SI{5}{\mega \hertz}. After filtering out the excitation wavelength, the luminescence of a single perovskite nanocube is collected by an inverted confocal microscope and sent to the optical characterization part. All the measurements were performed at room temperature.

\begin{figure}[t]
    \centering
    \includegraphics[width=\linewidth]{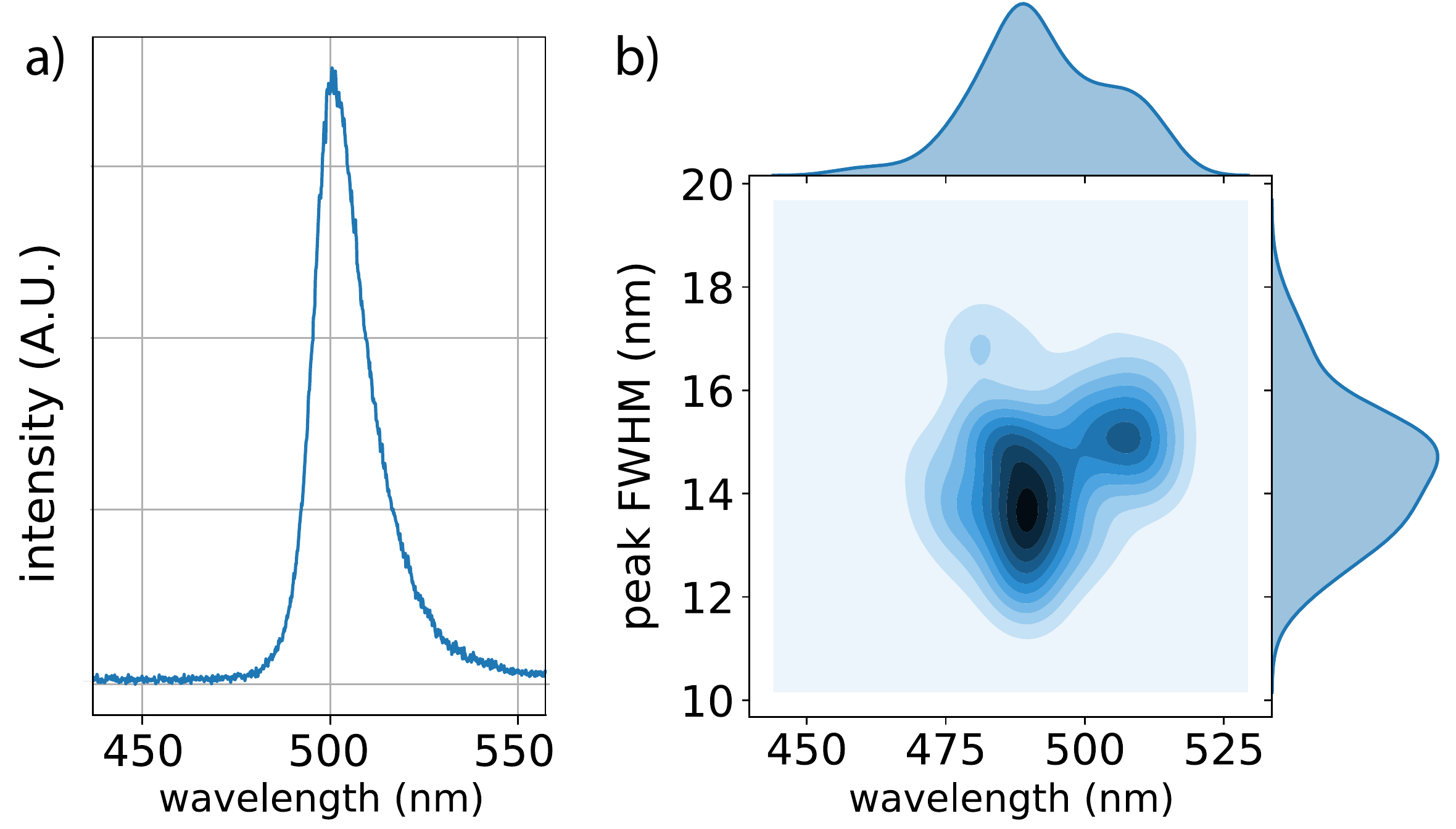}
     \caption{
         a)~Typical emission spectrum of a single \ncs{} and
         b)~central wavelength  and FWHM distributions of the emission for a sample of 24~emitters.
         }
  
    \label{fig:spettri}
\end{figure}

A typical emission spectrum is reported in Figure~\ref{fig:spettri}a and shows a central wavelength of \SI{500}{\nano \meter} with a full width at half maximum (FWHM) of about \SI{15}{\nano \meter}. The distribution of the emitted wavelengths with the corresponding FWHM is shown in Figure~\ref{fig:spettri}b for 78~emitters. 

Here, we clearly see two main peaks with central emission wavelength (CEW) at \SI{485}{\nano \meter} and \SI{510}{\nano \meter} respectively. The mean FWHM of the emitted light is about \SI{14}{nm} for the first peak and \SI{15}{nm} for the second one.
By comparison with the bulk perovskite emission wavelength at about \SI{520}{\nano \meter}\cite{mannino2020temperature,liashenko2019electronic,ng2018tunable}, we attribute the former emission to small single photon emitters \ncs{} with slight quantum confinement and the latter emission to large size nanocubes for which the confinement is mostly absent. As shown in the  following,  the small size nanocubes exhibit strongly antibunched emission, highlighting for the first time in this kind of emitters the crucial role of the charge confinement on their quantum properties.
We report a study over 24 different emitters.
For each emitter we measured the emitted power as a function of the excitation intensity. The background is subtracted from the experimental data. By fitting the data using the following saturation function, we are able to extract the saturation intensity $I_{sat}$:
\begin{equation}
\label{eq:saturation}
    P=A\cdot \bigg[ 1- e^{-\frac{I}{I_{sat}}}
\bigg] + B \cdot  I
\end{equation}
The first term of the sum represents the saturating part due to the single exciton component while the second is due to bi-exciton emission\cite{park2011NearUnity,rainoUnderestimatedEffectPolymer2019,raja2016Encapsulation}. Specifically $I_{sat}$ is the saturation intensity while $A$ and $B$ depend on the intensity of single and bi-exciton components of the emission.

A typical saturation curve is shown in Figure~\ref{sch:setup1}b. 
To minimize the effect of the blinking on the data analysis, 
multiple measurements are taken for each experimental intensity in the graph,
and only the one with the strongest emission is kept. 
We observed a median $I_{sat}$ of \SI{0.56}{\watt / \micro \meter^2}. 
The presence of a bi-exciton component indicates that the correlation function $g^{(2)}\of{0}$ will depend on the intensity at which we excite the emitters.
We clearly observed this behavior when repeating the measurements for several excitation intensities. To perform reliable measurements on several emitters and to ensure a signal well above the noise level, we performed all the measurements at the saturation intensity.

\subsection{Effect of the sample dilution on the photobleaching}
Usually perovskite nanocrystals at room temperature suffer from fast photo-bleaching when they are exposed to light. Several studies presented lead halide perovkites \ncs{} emission instability\cite{huang2017Lead}. Often, the monitoring of the spectral stability over few minutes is used to evaluate the photo-stability\cite{raja2016Encapsulation,rainoUnderestimatedEffectPolymer2019}, showing that the CEW shifts to more than \SI{10}{nm} after few tens of seconds for perovkite nanocrystals directly deposited on a glass-plate.
Even when they are encapsulated with polystyrene, the longest measurement time recorded in the literature by~\citeauthor{rainoUnderestimatedEffectPolymer2019}\cite{rainoUnderestimatedEffectPolymer2019} was around \SI{100}{\second}. The observed spectral drift has been attributed to the degradation of the nanocrystals, resulting in a progressive reduction of their sizes.
In our case, the study of the photo-stability of the sample synthesized with the procedure described above shows a significant improvement of the photostability of the emitters under illumination. 
\begin{figure}[t]
    \centering
    \includegraphics[width=\linewidth]{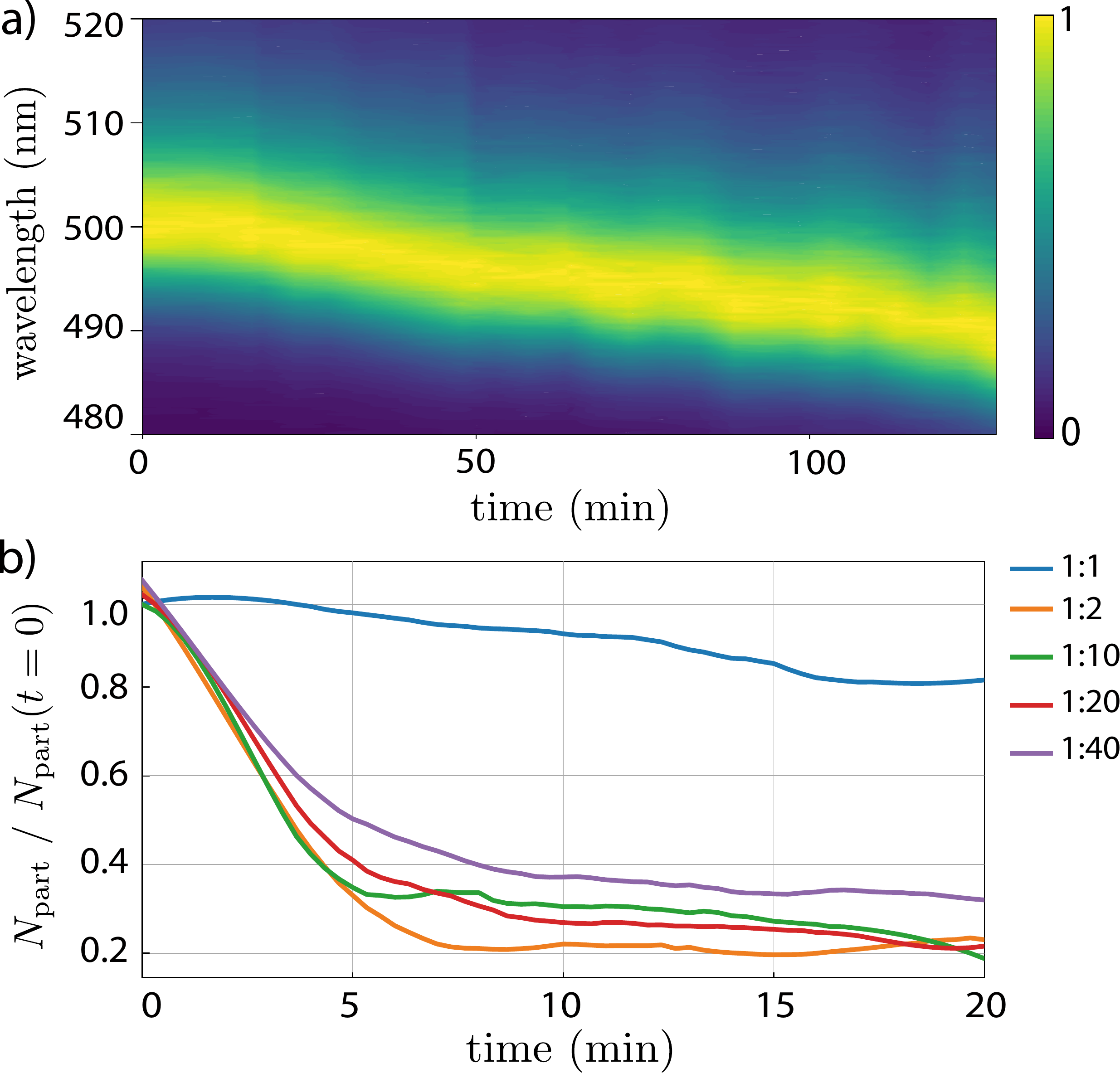}
     \caption{ Robustness of perovskites \ncs{}. 
        a)~Evolution vs time of the emission wavelength of a single perovskite \nc{} excited at the saturation intensity. Each spectrum is normalized.
        b)~Percentage of \ncs{} still emitting after a certain time under strong illumination as a function of the dilution of the original solution: 1:1, 1:2, 1:10, 1:20, 1:40 where 1:X means x times dilution.
        }
    \label{fig:robust}
\end{figure}
We prepared six samples using different dilutions starting from the most concentrated solution (typically with a molar concentration between \SI{1}{\micro M}-\SI{10}{\micro M}), up to a dilution of 1:100, using toluene as solvent.
We started by analyzing the highest concentration sample. With this concentration value, we are still able to individually address each emitter and to collect its luminescence for more than \SI{1}{\hour} with strongly reduced bleaching effects. This can be clearly see in Figure~\ref{fig:robust}a: an emitter is excited for \SI{2}{\hour} while its emission spectrum is collected every \SI{5}{\minute}. The figure reports the evolution of the normalized emission spectra versus time. A blue-shift is observed, as already reported by \citeauthor{rainoUnderestimatedEffectPolymer2019}\cite{rainoUnderestimatedEffectPolymer2019}, but on a much longer time scale: indeed our emitters exhibit a blue-shift of less than \SI{10}{\nano \meter} after two hours, showing a remarkable stability, two orders of magnitude better than previously observed\cite{parkRoomTemperatureSinglePhoton2015,rainoUnderestimatedEffectPolymer2019}.
To the best of our knowledge, it is one of the most robust samples reported in the literature.

This robustness is strongly related to the concentration of the emitters in the solution, and drops fast when we dilute the sample. 
Our final objective is to couple a single emitter to a tapered optical nanofiber to develop an integrated singe photon source. For this, we need to use a strongly diluted sample. It is then crucial to investigate the behavior of the emitters as a function of the concentration.

The results of a systematic study of the effect of the dilution are reported in Figure~\ref{fig:robust}b where
each sample is strongly illuminated with light from the LED lamp while a video of the sample emission is recorded. Analyzing each frame of the video (we take a frame each 20 seconds) we can estimate the number of \ncs{} that are still emitting from the first frame. We can clearly see that only the concentrated sample (so-called (1:1)) lasts for a long time with more than $80\%$ of the emitters still working after \SI{20}{\min}, while for all the other samples (from dilutions (1:2) to (1:100)) half of the emitters have bleached after \SI{3}{\min}.
We attribute this effect to the dynamic bonding of the ligand to the perovskite nanocrystal surface. Under dilute conditions, the free ligands can hardly find the surface of another nanocrystal. This displaces the equilibrium between bound and unbound ligands towards the latter. As a result, the dilution process leads to poorly passivated nanocrystals which can easily bleach.

\subsection{Blinking characterization}

The emitted intensity of a NC is not constant in time but tends to fluctuate: this phenomenon is known as blinking. This behavior has been reported for several kinds of quantum emitters such as single molecules\cite{yip1998Classifying,dickson1997Blinking}, \ch{Si}~nanocrystals\cite{mason1998Luminescence} and \ch{CdSe}/\ch{CdS} colloidal quantum dots\cite{nirmal1996Fluorescence,shimizu2001Blinkingb, tang2005Single,peterson2009Modified,mahler2008Nonblinking,houel2015Autocorrelation}. 

\begin{figure}[ht]
    \centering
    \includegraphics[width=\linewidth]{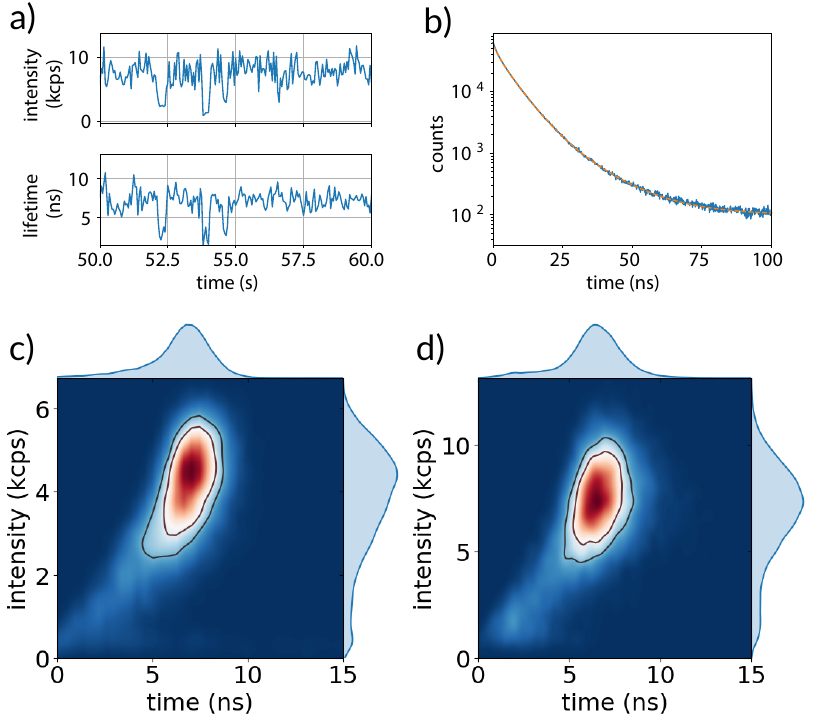}
   
    \caption{\label{fig:FLID}(a)~Blinking trace (upper graph) and lifetime (lower graph) of a single \nc{}.
    (b)~Lifetime of a single \nc{}, fitted with a triple-exponential decay model. We obtain \SI{1.4}{\nano \second}, \SI{6.1}{\nano\second} and \SI{14.7}{\nano\second}, corresponding respectively to the lifetimes of the biexciton, grey and neutral emission states.
    (c,d)~Fluorescence lifetime-intensity distribution (FLID) images of a single emitter, excited at half~(c) and twice~(d) the saturation intensity. The closed curves contains $50\%$ (inner one) and $68\%$ (outer one) of the occurrences.}
\end{figure}

Usually the fluorescence blinking is attributed to the trapping of charge carriers. In particular, it has been shown that we can distinguish two types of blinking: Type~A, in which the core is left effectively charged and the low fluorescence state is caused by the recombination due to the Auger effect, and Type~B, in which the trapped charge can recombine non-radiatively with its opposite charge and the blinking is due to fluctuations in the trapping rate.

A well established method to experimentally distinguish between Type~A and Type~B blinking is to study the spontaneous emission lifetime dependence on the emission intensity: in Type~A blinking the lifetime is expected to depend on the emission intensity\cite{galland2011two} while in Type~B the lifetime should not depend on the intensity\cite{galland2011two,brawand2015Surfacea}.

When the typical blinking time is too short with respect to the chosen binning of the time-trace curve, it becomes impossible to completely distinguish between grey and bright states: in this case it is more appropriate to describe the behavior of the emitter in terms of flickering \cite{galland2011two}. 

Our emitters, like the vast majority of perovskite \ncs{} reported in the literature\cite{trinh_organicinorganic_2018,seth_fluorescence_2016}, 
show a clear flickering behavior in their emission time trace. 
A zoom of a typical blinking trace (i.e.the emitted intensity as a function of time) of one emitter is shown in the upper box of Figure~\ref{fig:FLID}a. The signal is binned with a binning time of \SI{50}{ms}. The complete blinking trace is reported in the supplementary information. Interestingly, the inspection of this trace indicate a reduced blinking  with respect to the typical behavior of this kind of perovskites reported in literature \cite{parkRoomTemperatureSinglePhoton2015}.   
In the lower box of Figure~\ref{fig:FLID}a the mean lifetime versus time is reported: a clear correlation between the two curves is observable which indicates the presence of a type A blinking for these emitters.
 
In order to perform a more quantitative analysis, we fit the lifetime histogram with a triple-exponential decay model:
\begin{equation}
    A_1\cdot e^{-\frac{t-t_0}{\tau_1}} + 
    A_2 \cdot e^{-\frac{t-t_0}{\tau_2}} +
    A_3 \cdot e^{-\frac{t-t_0}{\tau_3}} + B
    \label{eq:lifetime}
\end{equation}
with three different lifetimes, $\tau_1$, $\tau_2$ and $\tau_3$, corresponding to the neutral, the charged, and the biexciton state emission respectively\cite{galland2011two}. 
In equation~\eqref{eq:lifetime} $t_0$ represents the pulse arrival time, while $A_i$ are the amplitudes of each decay component; B is an offset added to take into account the dark counts. A typical result of the fitting procedure is shown in Figure~\ref{fig:FLID}b, showing a good agreement with the experimental results.
The dependence of the lifetime on the emitted intensity is commonly studied \cite{galland2011two,parkRoomTemperatureSinglePhoton2015,hu2015DefectInduced}  by using the fluorescence lifetime-intensity distribution (FLID).
In Figure~\ref{fig:FLID}c and~\ref{fig:FLID}d two FLIDs images corresponding to the same \nc{} are shown, corresponding to an excitation intensity of $0.5I_{sat}$ and $2I_{sat}$ respectively, obtained with a bin size of \SI{50}{\milli \second}. The dark curves delimit an area corresponding respectively to 50\% (inner one) and 68\% (outer one) probability of emission. 
As opposed to previous reports\cite{parkRoomTemperatureSinglePhoton2015}, showing a predominance of the grey state emission for high excitation powers, with a significant decrease of the emission intensity, remarkably, our perovskite nanocubes remain in the same proportion of bright and grey state while excited up to $2I_{sat}$ and more, without any significant decrease of the emission intensity.

\subsection{Quantum properties}
To characterize the quantum emission and verify if our \ncs{} can be used as single photon emitters, we measured the autocorrelation  function $g^{(2)}$ using a Hanbury Brown and Twiss (HBT) setup (see Figure 2). The autocorrelation function is defined as follows:
$$
g^{(2)}\left( \tau \right)= \dfrac{ \left<I(t) I(t+\tau)\right>}{\left< I(t) \right>^2},
$$ 
where $I$ is the intensity of the emission, $t$ the time and $\tau$ the delay between two different photon arrivals.

\begin{figure}[tbh]
    \centering
    \includegraphics[width=\linewidth]{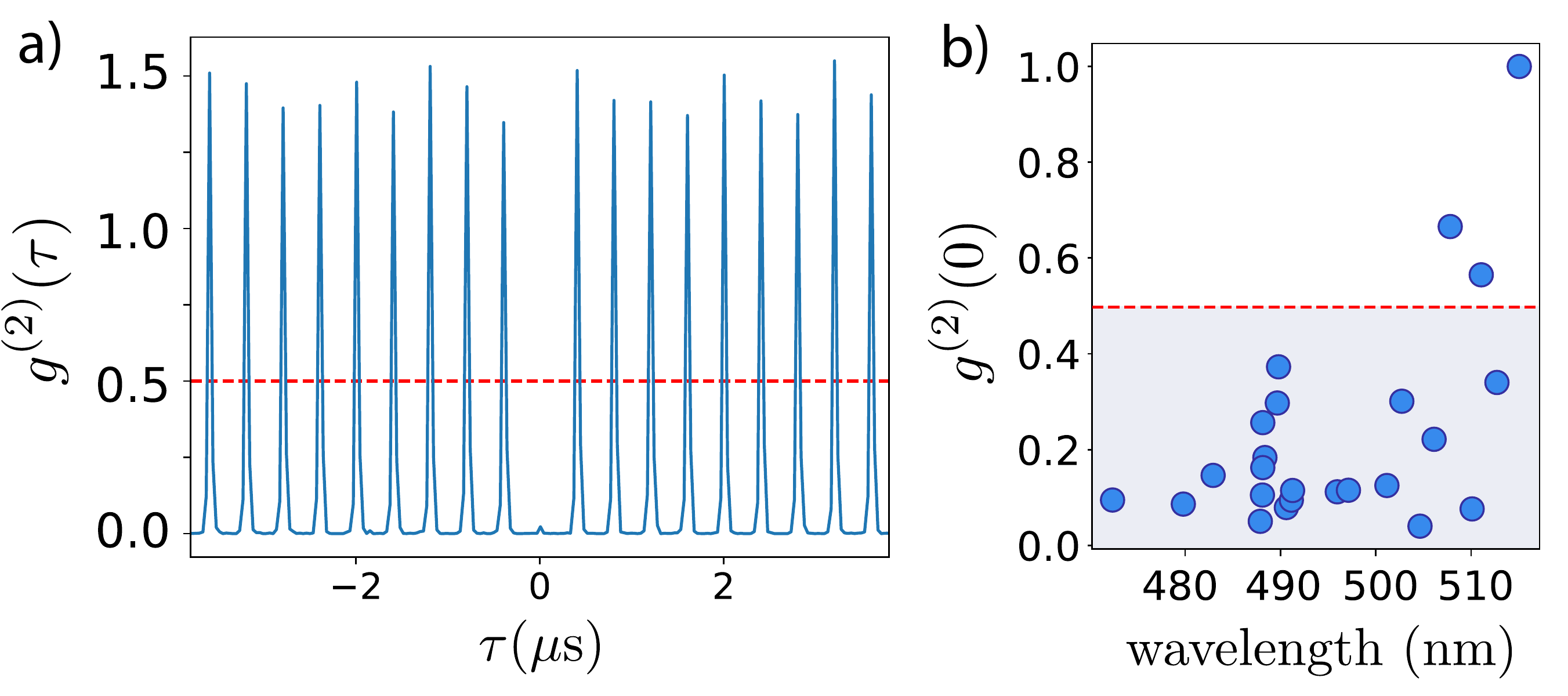}
    \caption{
        a)~Example of $g^{(2)}$ function of a single \nc{} emitting high quality single photons, measured with a repetition rate of \SI{2.5}{\mega \hertz}
        b)~Measured $g^{(2)}(0)$ values as a function of the central wavelength of the emission of the \nc{}. All the emitters were excited at their saturation intensity.
        }
    \label{fig:g2}  
\end{figure}
Experimentally the beam is divided in two parts by a 50/50 beam splitter (see Figure~\ref{sch:setup1}a) and sent to two avalanche photodiode single photon detectors (APDs). 
We record the arrival times with the time tagged time-resolved\cite{wahlTime} method. We then create a histogram of the relative arrival times of the photons with a time bin of \SI{18}{\nano \second},
shown in figure~\ref{fig:g2}a. 
Due to the blinking effect, the peaks close to the $0$ delay peak are higher than $1$\cite{manceau_cdsecds_2018}; on the other hand the peaks of the autocorrelation function tend to 1 for large delays compared to the characteristics blinking time. The histogram is thus normalized by setting the mean height of the peaks with $\tau\approx \SI{10}{\micro \second}$ to be $1$. This procedure is well established and documented in literature\cite{pierini2020Hybrid,manceau2014effect}. 
The background, mainly given by the dark counts of the APDs, is subtracted following the procedure described in supplementary materials.

The $g^{(2)}$ at $\tau=0$ delay is  $0.02$ according to Figure~\ref{fig:g2}a, well below $0.5$ showing clear signature of a single photon emission. We found that $50\%$ of our emitters show a very good single photon emission, with $g^{(2)}(\tau)<0.2$.

We perform a statistical analysis of the $g^{(2)}(0)$ as a function of the central emission wavelength, shown in Figure~\ref{fig:g2}b. We observe a clear degradation of the single photon emission for longer wavelengths. In particular, for CEWs above \SI{505}{\nano \meter}, the autocorrelation function value is always higher than~$0.5$, indicating that the single photon emission is lost. The reason for that is that for the biggest \ncs{} the quantum confinement is no longer effective and this has the effect to increase the emission wavelength and to reduce the quality of the single photon emission. 

\section{Nanofiber integration}
In this section, we describe the first realization of the coupling of a single perovskite nanocube with an optical taperd nanofiber, constituting a prototype of an interesting hybrid nanophotonic device for quantum technologies-oriented applications.
Tapered nanofibers are photonic waveguides obtained by stretching a
standard optical fiber while heating it, in order to reduce its diameter 
to some hundreds of nanometers. This results in a strong evanescent
field in the vicinity of the fiber\cite{kien2004Field}, which enables the
coupling of the light emitted by a nano-emitter located nearby directly into the nanofiber thus obtaining a compact and integrated single photon source\cite{joos2019complete}. This approach has been demonstrated with single atoms\cite{le2005spontaneous,klimov2004spontaneous} as well as single colloidal quantum dots\cite{fujiwara2011highly,yalla2012fluorescence,joos2018polarization} or with nanodiamonds containing single $\textrm{NV}^-$ defects\cite{vorobyovCouplingSingleNV2016}. This is of great interest for quantum applications, where nanofiber-based systems are rapidly developing\cite{nayak2018nanofiber}.

In order to couple our \ncs{} to the nanofiber, we place on top of it a \SI{20}{\micro \liter} droplet with a dilution 1:100 of the original solution in toluene using a micropipette. We then use a translation stage to carefully approach and eventually touch the nanofiber with the droplet, while monitoring the movement with a microscope. This is a critical step, as the nanofiber can easily break. When successful, this procedure results in several emitters deposited onto the nanofiber.

\begin{figure}[htb]
    \centering
    \includegraphics[width=\linewidth]{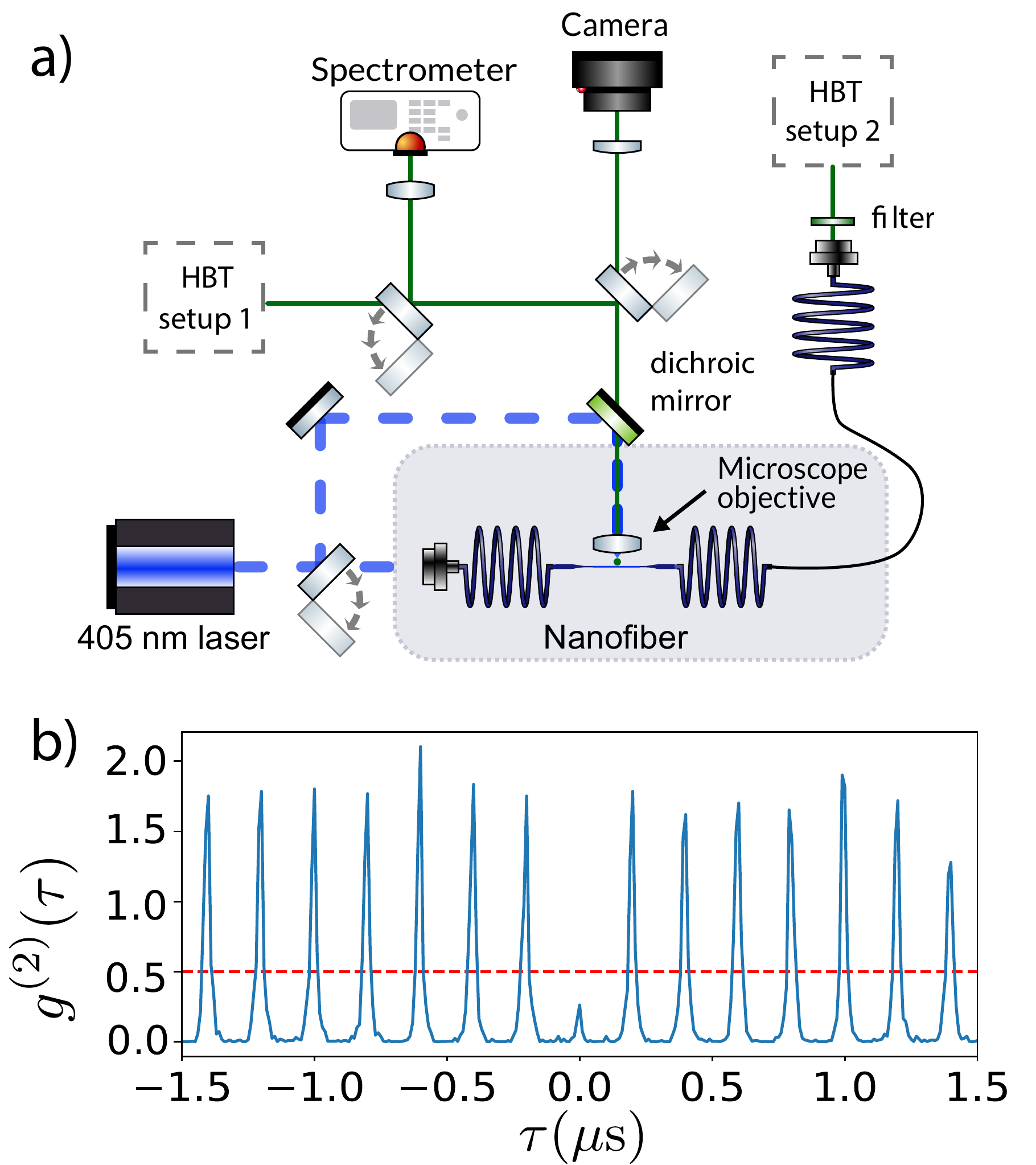}
    \caption{a) Setup used to measure single photons emission from a \nc{} coupled to the nanofiber.
    b) $g^{(2)}$ function of a single \nc{} emitting single photons into the nanofiber (HBT  setup 2 of figure a), measured with a repetition rate of \SI{5}{\mega \hertz}. 
    }
    \label{sch:setup_fiber}
\end{figure}

The setup used for this study is shown in Figure~\ref{sch:setup_fiber}.
We firstly select the emitter sending the laser in the fiber and detect the photoluminescence from the microscope objective.
Then, we excite a single \nc{} located on top of the nanofiber with a laser sent through that microscope objective. 
A significant fraction of the emitted light is coupled into the nanofiber and propagates to its end. The end of the fiber is plugged to an HBT set-up for a $g^{(2)}$ measurement following the procedure previously described. Optionally, photon antibunching and spectral measurements can be performed also in free-space via the light collected with the  microscope objective.  A typical antibunching measurement performed on the light collected via the nanofiber is shown in Figure~\ref{sch:setup_fiber}.
We are still able to detect single photon emission, and this demonstrates that a single perovskite \nc{} is coupled to a nanofiber and emits single photons directly inside it. 
These results show that our \ncs{} are not damaged by the deposition process and validate this approach for this kind of sources, that, due to the dilution, can be used for some minutes once deposited over the nanofiber.
Due to the versatility of the perovskite nanocubes and to  their emission of single photons with long coherence time\cite{utzat2019Coherent}, this technique opens the way to the realization of nanofiber-based, compact, integrated hybrid devices for indistinguishable single photon generation with solid state emitters.
In order to achieve this final goal, it is crucial to further improve the stability of the emitters thanks to a deeper understanding of the role that ligands and dilution play on their robustness.

\section{Conclusions}
We investigated the properties of highly-stable \ch{CsPbBr3} nanocrystals as single photon emitters. The effect of the dilution on the photo-stability was investigated, confirming a degradation when increasing the dilution. To further improve the stability, a promising approach, currently under investigation, is based on a better ligand control in order to increase the dilution without breaking the equilibrium of the colloidal solution.

A full characterization of the optical properties of these emitters was performed, highlighting the role of the charge confinement in their antibunching behaviour. A deep analysis of the blinking and single photon emission of perovskite nanocubes were presented, showing a strongly reduced blinking and a remarkable stability of the bright state emission as a function of the excitation power. This feature guarantees a very low $g^{(2)}(0)$ also for high excitation power.
Moreover, for the first time with such emitters, we have shown the coupling of a single perovskite nanocube with a tapered optical nanofiber. Thanks to the near-field interaction, single photons are emitted in the near field of the nanofiber demonstrating the proof of principle of a compact, integrated single photon source.
The coupling to other platforms, such as the ion-integrated waveguides one\cite{madrigal2016hybrid}, is also envisioned to obtain integrated single photon sources for quantum photonic applications.

\section{Experimental}
\subsubsection{Chemicals}
\ch{PbBr2} (Alfa Aesar, $98.5\%$), \ch{Cs2CO3} (Alfa aesar, $99,99\%$), oleylamine (OLA, Acros, $80-90\%$), oleic acid (OA, Sigma-Aldrich), octylamine (Oct.Am, Alfa aesar, $99\%$), Octanoic acid (Oct.Acid, Acros, $99\%$), octadecene (ODE, Acros Organics, $90\%$), toluene (VWR, rectapur).

\subsubsection{Caesium oleate precursor}
We mix in a \SI{50}{\milli \liter} three neck flask, \SI{350}{mg} of \ch{Cs2CO3} in \SI{20}{mL} of 
ODE and \SI{1.25}{mL} of OA. 
The content of the flask is stirred and degased under vacuum at room temperature for
\SI{25}{min}. The flask is heated at \SI{110}{\celsius} for \SI{15}{min}. The atmosphere 
is switched to nitrogen and the temperature raised to \SI{150}{\celsius}. The reaction is
carried on for \SI{15}{min}. At this stage the salt is fully dissolved. The temperature is 
cooled down below \SI{100}{\celsius} and the falsk degased under vacuum. Finally this solution 
is used as a stock solution.

\subsubsection{Nanocrystal synthesis}
 In a three neck flask, \SI{174}{mg} of \ch{PbBr2} is mixed in \SI{10}{mL} of ODE. 
 The flask is degased under vacuum at room temperature for \SI{15}{min}. Then, the 
 temperature is raised to \SI{120}{\celsius}. At \SI{105}{\celsius}, \SI{0.25}{mL} 
 of OLA is injected. Once vacuum has recovered, \SI{1.1}{mL} of OA is injected. 
 After 5 min, \SI{0.75}{mL} of oct.Am is injected. Once the vacuum has 
 recovered, \SI{0.75}{mL} of oct.Acid is injected. The solution is colorless at 
 this stage. The solution is further degased at \SI{120}{\celsius} for \SI{30}{min}. 
 The atmosphere is switched to nitrogen and the temperature raised to \SI{150}{\celsius}. 
 Around \SI{0.1}{mL} of CsOA solution is injected, and the solution turns turbid. 
 The solution is conducted for \SI{35}{min} and finally quickly cooled down by removing 
 the heating mantle and using a water bath. The solution is transferred to plastic tube and
 centrifuged. The supernatant is discarded. The pellet is dispersed in \SI{5}{mL} of toluene.
 The solution is centrifuged again at low speed (\SI{2000}{rpm} for \SI{1}{min}). The pellet is
 discarded and the colloidally stable supernatant is stored.
 
 \subsubsection{Material characterization}
 UV-visible absorption spectrum are obtained by diluting the nanoparticles in hexane and using a
 JASCO~V730 spectrometer. For transmission electron microscopy, a dilute solution of nanocrystals is drop-casted
 onto a copper grid coated with a thin amorphous carbon layer. The grid is then degased under
 secondary vacuum overnight. Imaging was conducted with a JEOL~2010 microscope operated at
 \SI{200}{kV}.

\begin{acknowledgement}

The project is supported by ERC starting grant blackQD (grant n° 756225), by ANR grants IPER-Nano2 (ANR-18CE30-0023-01), Copin (ANR-19-CE24-0022), Frontal (ANR-19-CE09-0017), Graskop (ANR-19-CE09-0026) and by the European Union’s Horizon 2020 research and innovation program under grant agreement No 828972 -Nanobright. AB and QG are members of the Institut Universitaire de France (IUF). The authors thank J.-P. Hermier for inspiring discussions.

\end{acknowledgement}

\begin{suppinfo}

Supporting information available: Integral fluorescence trace, procedure for FLID generation, noise cleaning procedure for $g^{(2)}(\tau)$ measurements, procedure for \ncs{} deposition on nanofibers.
This material is available free of charge via the internet at \url{http://pubs.acs.org}

\end{suppinfo}

\bibliography{achemso-demo}

\end{document}


\section{Fluorescence Trace}
In Figure~\ref{fig:timetrace_1d}, the timetrace of a single \nc{} is shown. The measurements lasted \SI{15}{\minute} and we did not observe any visible bleaching during this time. The emitter was used for other kind of measurements after this one, and lasted for more than one hour.
\begin{figure}[H]
    \centering
    \includegraphics[width=\textwidth]{Pictures/em1d_timetrace.pdf}
    \caption{Timetrace of a single \nc{}, performed with a binning of \SI{50}{\micro \second}.}
    \label{fig:timetrace_1d}
\end{figure}

\section{Generation of FLID images}
The FLID images shown in figure~4 are generated following these different steps:
\begin{enumerate}
    \item The luminescence trace is generated using a binning time of \SI{50}{\nano\second} as shown in Figure~4a of the main paper.
    \item For the photons arriving in each bin, the mean arrival time is computed, making sure to previously select only the photons arrived in less then \SI{50}{ns} from the peak: this reduces the effect of the background in bins with low intensity emission.
    \item For each bin, given the mean arrival time calculated before and the mean intensity, a point can be placed in the intensity-time space.
    \item The kernel density estimation\cite{rosenblatt1956Remarks,parzen1962Estimation} is used to obtain the FLIDs shown in Figure~4c and~4d of the main paper.
\end{enumerate}

\section{Noise cleaning on $g^{(2)}(\tau)$ measurements}
To correctly perform the $g^{(2)}(\tau)$ measurement we need to firstly made some considerations on how the coincidences are counted. We call $M(\tau)$ the number of counts measured at a given delay $\tau$. Each count can be generated either by a start from the signal or from the background and by a stop from the signal or from the background. Calling $s(\tau)$ the probability to have a start (or stop) generated by the signal and $b(\tau)$ the probability to have the start (or stop) generated by the background, we can write (when both $b(\tau)$ and $c(\tau) \ll 1$): 
\begin{equation}
    M(\tau)=C\tonda{b(\tau)+s(\tau)}\tonda{b(\tau)+s(\tau)}
\end{equation}
where C is a constant of proportionality. We call $M'=\frac{M}{C}$.
We consider $\tau_b$ in between two consecutive peaks, we know that there is no signal there and $s(\tau_b)=0$.
\begin{equation}
    M'(\tau_b)=b^2(\tau)
\end{equation}
To find $M_{c}(\tau)=Cs^2(\tau)$ we solve the system of these two equations and  we find that 
\begin{equation}
    M_{c}(\tau)=M(\tau)+M(\tau_b)-2\sqrt{M(\tau)}\sqrt{M(\tau_b)}
\end{equation}
With this formula we obtained the $g^{(2)}(\tau)$ histogram cleaned from the background counts.

\section{\ncs{} deposition on nanofibers}
After fabrication of the nanofiber, a particular care is needed for the \ncs{} deposition: the nanofiber is very sensitive to any mechanical solicitation and it is very easy to break it. In addition, any dust deposited over it can perturb the surrounding evanescent field and degrade the nanofiber transmission:
the whole procedure needs thus to be performed under a clean laminar flux, to avoid any possible contamination.
We create a droplet of \SI{20}{\micro \liter} with the solution containing the \ncs{} at the point of a micropipette, and gently approach the fiber using a translation stage. A scheme of this procedure is shown in figure~\ref{fig:micripipette}.
\begin{figure}
    \centering
    \includegraphics[width=\linewidth]{Pictures/SchemeFiber_all.pdf}
    \caption{Scheme of the procedure for deposition of single \ncs{} over a nanofiber. A droplet of \SI{20}{\micro \liter} is created (a), and the fiber is touched with it (b) using a precise translation stage. The droplet is removed and this results in one or more \ncs{} deposited on the fiber (c).}
    \label{fig:micripipette}
\end{figure}
We monitor the procedure with the microscope objective shown in Figure~6 of the main article. When the droplet comes in contact with the fiber, we gently move the droplet away from it. This procedure often results in one or more emitters deposited over the fiber: this can easily be verified by shining the excitation laser inside the fiber and filtering out  the excitation beam to collect only the perovskite nanocube emission wavelength.
\bibliography{achemso-demo}